 \documentclass[a4paper,11pt]{article}
 \pdfoutput=1 
 
 \usepackage{jheppub} 
 
 \usepackage[T1]{fontenc} 
 \usepackage{float}
 \usepackage{epstopdf}
 \usepackage{amsmath}
\makeatletter
\gdef\@fpheader{}
\makeatother

\title{\boldmath Energy-dependent topological black holes in massive gravity coupled to double-logarithmic electrodynamics}

\author[1,2]{Askar Ali} 
\affiliation[1]{Department of Mathematics, Quaid-i-Azam University, Islamabad, Pakistan}

\affiliation[2]{Department of Sciences and Humanities, National University of Computer and Emerging Sciences, Peshawar 25000, Pakistan}

\emailAdd{askarali@math.qau.edu.pk}

\abstract{In this paper, I investigate the nonlinear magnetized black holes of massive gravity's rainbow. In order to achieve this, I use the double-logarithmic electromagnetic field as a matter source and derive the new anti-de Sitter black hole solution. The effects of the massive graviton, rainbow functions, and nonlinearity of the electromagnetic field on the black hole's inner and outer horizons are also studied. The black hole's thermodynamic properties are studied, and the mathematical expressions for temperature, heat capacity, and Gibbs free energy are found. The extended first law and Smarr's formula are derived as well. Using the behaviour of thermodynamic quantities as a starting point, I also study the local and global thermodynamic stabilities.
\vspace{80 mm}
}

\notoc 

\begin{document}
\maketitle
\flushbottom 


\section{Introduction}
\label{sec:intro}
Since Einstein's general relativity (EGR) produces many important predictions, however, it requires corrections when it comes to high energy regimes. Different attempts have been worked out for it at a fundamental level and modified gravity theories such as Lovelock gravity \cite{1,2}, scalar-tensor theories \cite{3,4}, brane world cosmology \cite{5,6} and $f(R)$ gravity \cite{7,8,9,10} are constructed. In addition to these theories, another modification of EGR called doubly general relativity or gravity's rainbow \cite{11,12} has also been developed. It is expected that gravity's rainbow may be one of the successful candidates which explain different aspects involving ultraviolet (UV) divergences.
 
To understand the framework of gravity's rainbow, one should need to review the model of doubly special relativity as it gives the historical origin for it. It is very familiar that Einstein's special theory of relativity was constructed on the basis of two postulates \cite{13}; (i) the laws of physics take the same form in all inertial frames, and (ii) the constancy of the speed of light in a vacuum for any observer, regardless of the relative motion of the light source. Along with these two postulates, if one assumes an upper bound on the energy associated with a test particle then doubly special relativity is introduced \cite{14,15,16}. This implies that the energy corresponding to the test particle should not exceed the Planck energy. It is obvious that the generalization of Einstein's special theory of relativity is EGR. Therefore, under the effects of the high energy regime, the generalization to doubly special relativity leads to the so-called doubly general relativity or gravity's rainbow \cite{12}. 

One more way to develop the model of gravity's rainbow is to generalize the Einstein's energy-momentum relation. Hence, one may write 
\begin{equation}
E^2h^{2}(\varepsilon)-p^2g^{2}(\varepsilon)=m^2\label{1q}.
\end{equation}
The parameter $E$ represents the test particle's energy whereas $E_p$ denotes the Planck energy. The relationship between these energies is described by the condition $E/E_p=\varepsilon\leq1$. This implies that $E$ can never attain a greater value than the Planck energy \cite{17}. Note that, the functions $g(\varepsilon)$ and $h(\varepsilon)$ are called the rainbow functions \cite{18}. It is worth noting that the rainbow functions should need to satisfy the infrared (IR) limiting conditions: $g(\varepsilon)\rightarrow1$ and $h(\varepsilon)\rightarrow1$ in the limit $\varepsilon\rightarrow0$. This restriction assures us that EGR is a specific case of the gravity's rainbow.  Recently, the model of gravity's rainbow becomes very interesting due to its implementation in the subject of modern theoretical physics. The problem of information paradox has also been investigated within this formulation of gravity \cite{19,20}. In addition to this, it also explains the well-known uncertainty principle \cite{21,22} and the black hole's remnants due to quantum evaporation \cite{23}. In the background of cosmology, it is expected that the model of gravity's rainbow may solve the issues that arises due to the big-bang singularity \cite{24,25}. The issues caused by the initial singularity \cite{26} and the stability in Einstein static universe have also been investigated in this theory \cite{27}. It has also been observed that the formula of mass associated with the white dwarfs and neutron stars is monotonically increasing function of $g(\varepsilon)$ and $h(\varepsilon)$ \cite{28,29,30}. Moreover, the relationship of gravity's rainbow with Horava-Lifshitz gravity can be established from the suitable scaling of energy functions \cite{31}. Physical aspects of massive objects such as black hole thermodynamics in gravity's rainbow have also been studied \cite{32,33}. Similarly, the effects of electromagnetic fields on the thermodynamic properties of these objects were examined in Ref. \cite{34}. Furthermore, the effects of rainbow functions on the thermodynamic stability of black holes in modified gravities were studied in Refs. \cite{36,37,38,39}. 

In the field of astrophysics and cosmology, the experimental verifications of EGR motivate one to consider it a successful theory describing gravitational phenomena. The assumption of a constant $\Lambda$ in the Einstein-Hilbert Lagrangian leads one to the results produced by Einstein-$\Lambda$ gravity with dark energy prediction. On the other side, the model of EGR is also consistent with the interactions of massless gravitons having spin 2. Besides its consistency, the associated quantum theory in the presence of massless gravitons is completely non-renormalizable \cite{1a}. Thus, it would be quite motivating to study gravitational fields in the background of massive gravity. The model of massive gravity with a massive spin 2 particle propagations can be developed if one includes the massive graviton terms in the Einstein-Hilbert action. These additional contributions imply that the graviton has a nonzero mass denoted by $m_g$. Note that, one would obtain the action function of EGR for $m_g=0$. The investigation of massive gravities in both flat and curved geometries which results in the absence \cite{2a}, and appearance of ghosts \cite{3a} has also been performed. The quantum mechanical aspects corresponding to linear as well as nonlinear massive gravities in the presence of ghost-free field \cite{4a} were investigated in Refs. \cite{5a, 6a}. Additionally, massive gravity theories (for example, some outstanding reviews in \cite{7a, 8a}) have received much attention due to their several interesting consequences, for instance, analysis of the recent observations related to dark matter \cite{9a, 10a}. It is also shown that our Universe expands with positive acceleration where no influence from the dark energy sector \cite{11a, 12a} has been arisen. In this paper, I consider the contributions of the nontrivial anti-de Sitter massive theory \cite{13a, 14a} in the background of energy-dependent spacetime and investigate the charged black holes within this setup. The motivation behind this comes from the fact that the behaviour of graviton is matching with that of a lattice in the case of a holographic conductor \cite{15a}. Another interesting phenomenon that supports the model of massive gravity is the metal-insulator transition \cite{16a}.

Massive gravity not only requires the dynamical metric $g_{\mu\nu}$ but it needs the fiducial reference metric $f_{\mu\nu}$ and non-derivative potentials denoted by $U_i$ as well. It is very interesting that associated with any suitable choice of $f_{\mu\nu}$ a unique family of massive gravities \cite{17a} can be constructed. The dRGT massive theory comes out to be ghost-free for many reference metrics, i.e., the Minkowskian and degenerate (singular) reference metrics \cite{18a,19a,20a,21a,22a}. Investigations of different cosmological phenomena showed that they are mainly relying on the choices of Minkowskian reference metrics \cite{21a,22a,23a,24a,25a}. Recently, many exact black hole solutions have been derived in the background of massive gravity \cite{13a,26a,27a,28a,29a,30a}. Among them, the anti-de Sitter black holes are important in different applications of gauge/gravity duality \cite{13a} and black hole chemistry \cite{31a}. 

In this work, I am interested in finding the new black hole solution of the massive gravity's rainbow sourced by nonlinear electrodynamics (NLED). Since the last century, nonlinear electromagnetic fields acquired much interest and several models were formulated for their description. The most famous model that describes nonlinear electromagnetic fields is the Born-Infeld (BI) electrodynamics \cite{81}. The BI action was also emerged in the super-string theory's low energy effective action \cite{82,83}. In contrast to Maxwell's electrodynamics, this model of NLED enables one to calculate the well-defined electric field at the origin. Besides the Born-Infeld model, many other models describing nonlinear electromagnetic phenomena were also introduced in the literature \cite{84,85,86,87,88}. Even though the NLED was developed for removing the divergences from Maxwell's theory, it also becomes very important in the study of charged black holes and their thermodynamical properties. The exact static black hole solution in EGR with NLED source was derived in Ref. \cite{89}. The Lagrangian density describing this NLED source defines a large family of nonlinear theories which contains Born-Infeld and Euler-Heisenberg models \cite{90} as specific cases. Recently, some new formulations regarding NLED have also been considered as sources of gravity \cite{91,92,93,94,95}. The solutions obtained as a result of these models characterize the regular black holes and asymptotically behave like the Reisnner-Nordstr\"{o}m (RN) solution. Furthermore, the effects of the NLED models on the black holes of various gravity theories were extensively investigated in Refs. \cite{98,99,100,102,103,107,108,109,113,114,116,117,118,119,120,121,122,123,124,125,126,126b,127,127a,127b,127c,127d,127f,127f1,127f2,127g,127h,127i,127j,127k,127r1,127r2}. Among the earlier mentioned work, the investigation of charged black holes in massive gravity has been done in Refs. \cite{127a,127b,127c,127d,127f,127f1,127f2}. Rotating black branes in higher curvature gravities within the model of NLED were studied in Refs. \cite{128a,my1}. Furthermore, dilatonic black holes were studied in Refs. \cite{128b,128c,128d,128E}. Without including dark energy, NLED is also significant for studying the inflation period during the start of the Universe \cite{128,129}. Some models of NLED were also used for the investigation regarding the expansion of the Universe \cite{130,131,132,133,134,135}. 

  In this paper, the nonlinear double-logarithmic electrodynamics \cite{88} is coupled to massive gravity's rainbow and the physical properties of new magnetized black holes are studied. It is worth noting that the model of double-logarithmic electrodynamics depends on both Maxwell invariants $\digamma=F_{\mu\nu}F^{\mu\nu}$ and $\Upsilon=F_{\mu\nu}\tilde{F}^{\mu\nu}$. It is shown that this model of NLED produces unavoidable effects on the black hole's thermodynamic stability and the magnetized black hole solutions in both four and higher dimensions acquired the same behaviour as those solutions which were obtained with Born-Infeld sources \cite{127j,127k}. Using this NLED source, the solution describing energy-dependent topological black holes and the exact expressions of different thermodynamic quantities are derived. Associated with these quantities, the generalized first law and Smarr's relation in the present scenario of massive gravity's rainbow are also constructed \cite{138,139,141,143}.

The present paper is ordered as. I coupled massive gravity with double-logarithmic electrodynamics and derive the modified gravitational field equations from the variational principle in Section 2. The effects of gravity's rainbow are also taken into account and a class of new black hole solutions is find out. The thermodynamic characteristics of these black holes are covered in Section 3. At last, the paper is ended with some concluding remarks in Section 4.

\section{Energy-dependent magnetized black hole solution} 

Here, the $d$-dimensional black hole solution with various horizon topologies is derived. To include the effects of gravity's rainbow, one can consider a one-parameter family of orthonormal frame fields that leads to the creation of metrics defined through the relation $g^{\mu\nu}(\varepsilon)=e^{\mu}_a(\varepsilon)e^{a\nu}(\varepsilon)$ such that $\varepsilon=E/E_p$ refers to the ratio of test particle and Planck energies. In this setup, the following spherically symmetric and an energy-dependent ansatz is introduced
 \begin{equation}
 ds^2=-\frac{f(r)}{h^2(\varepsilon)}dt^2+\frac{1}{g^2(\varepsilon)}\bigg(\frac{dr^2}{f(r)}+r^2d\Omega_k^2\bigg).
 \label{1}
 \end{equation}
 Here, $d\Omega^2_k$ stands for the metric associated with the $(d-2)$-dimensional hyper-surface. The constant curvature of this surface is given by $(d-3)(d-2)k$, while $V_{d-2}$ denotes its volume. One can express the explicit form of $d\Omega_k^2$ as
 \begin{equation}\begin{split}
 d\Omega_k^2=\left\{ \begin{array}{rcl}
 d\theta^2_1+\sum_{i=2}^{d-2}\prod_{j=1}^{i-1}\sin^2{\theta_j}d\theta_i^2, & \mbox{for}
 & k=1 \\\sum_{i=1}^{d-2} d\phi^2_i, & \mbox{for} & k=0 \\
 d\theta^2_1+\sinh^2{\theta_1}\big(d\theta^2_2+\sum_{i=3}^{d-2}\prod_{j=2}^{i-1}\sin^2{\theta_j}d\theta_i^2\big), & \mbox{for} & k=-1
 \end{array}\right.\label{2}
 \end{split}
 \end{equation}
 
The action of massive gravity's rainbow with matter sources can be defined as 
\begin{equation}
I=I_g+I_m,
\label{3}
\end{equation}
where $I_m$ is the action describing matter contents while $I_g$ describes the massive gravity's rainbow with cosmological constant $\Lambda$ as 
\begin{equation}
I_g=-\frac{1}{16\pi}\int d^dx\sqrt{-g}\big[R(\textbf{g},\varepsilon)+m^2_g(\varepsilon)\Sigma_{i=1}^{4}\alpha_iU_i(\textbf{g},\textbf{f})-2\Lambda(\varepsilon)\big].
\label{4}
\end{equation}
Here, $R(\textbf{g},\varepsilon)$ is the Ricci scalar, $\Lambda(\varepsilon)$ is the cosmological constant whereas $\textbf{g}$ and $\textbf{f}$ are the metric and fiducial metric tensors, respectively. Also, $m_g(\varepsilon)$ represents the graviton's mass, $\alpha_i$'s are the parameters of massive gravity, and $U_i$'s are the non-derivative potentials. It is worth noting that these potentials are symmetric polynomials corresponding to the eigenvalues of the $d\times d$ matrix $\chi^{\mu}_{\nu}=\sqrt{g^{\mu\alpha}f_{\alpha\nu}}$. One can find them as
\begin{eqnarray}\begin{split}
U_1=[\chi], U_2=[\chi]^2-[\chi^2], 
U_3=[\chi]^3-3[\chi][\chi^2]+2[\chi^3],\\
U_4=[\chi]^4-6[\chi^2][\chi]^2+8[\chi^3][\chi]+3[\chi^2]^2-6[\chi^4]. \label{5}\end{split}
\end{eqnarray}
After the variation of Eq. (\ref{3}) with respect to the metric $g_{\mu\nu}$, one can get the equations describing gravitational field as
\begin{equation}
R_{\mu\nu}-\frac{1}{2}g_{\mu\nu}R+\Lambda(\varepsilon) g_{\mu\nu}+m^2_g(\varepsilon)X_{\mu\nu}=\kappa T_{\mu\nu}.
\label{6}
\end{equation}
Here $T_{\mu\nu}$ represents the matter-tensor given by
\begin{equation}
T_{\mu\nu}=-\frac{2}{\sqrt{-g}}\frac{\delta I_m}{\delta g^{\mu\nu}},
\label{7a}
\end{equation}
and 
\begin{eqnarray}
\label{7}
\begin{split}
X_{\mu\nu}&=-\frac{\alpha_1}{2}\bigg(U_1g_{\mu\nu}-\chi_{\mu\nu}\bigg)-\frac{\alpha_2}{2}\bigg(U_2g_{\mu\nu}-2U_1\chi_{\mu\nu}+2\chi^2_{\mu\nu}\bigg)-\frac{\alpha_3}{2}\\&\times\bigg(U_3g_{\mu\nu}-3U_2\chi_{\mu\nu}+6U_1\chi^2_{\mu\nu}-6\chi^3_{\mu\nu}\bigg)-\frac{\alpha_4}{2}\bigg(U_4g_{\mu\nu}-4U_3\chi_{\mu\nu}\\&+2U_2\chi^2_{\mu\nu}-24U_1\chi^3_{\mu\nu}+24\chi^4_{\mu\nu}\bigg).
\end{split}
\end{eqnarray}

\begin{figure}[h]
    	\centering
    	\includegraphics[width=0.8\textwidth]{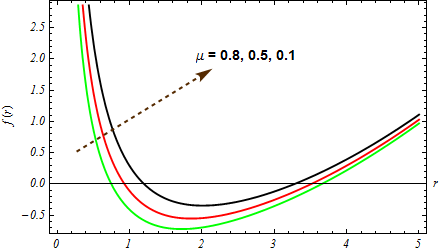}
    	\caption{Behaviour of the solution $f(r)$ (Eq. (\ref{17})) for $\Lambda(\varepsilon)=-0.3$, $Q=0.5$, $\beta(\varepsilon)=1.5$, $C=1$, $d=4$, $\alpha_1=1$, $\alpha_2=-1$, $\alpha_3=1$, $\alpha_4=-1$, $m_g(\varepsilon)=0.05$, $k=-1$, $g(\varepsilon)=1.1$ and $h(\varepsilon)=1.2$.}\label{skr1}
    \end{figure}

\begin{figure}[h]
    	\centering
    	\includegraphics[width=0.8\textwidth]{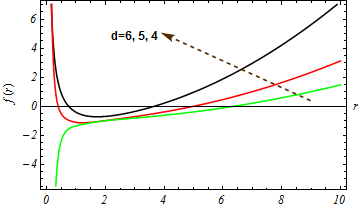}
    	\caption{Behaviour of the solution $f(r)$ (Eq. (\ref{17})) for $\Lambda(\varepsilon)=-0.3$, $Q=0.5$, $\beta(\varepsilon)=1.5$, $C=1$, $\mu(\varepsilon)=0.8$, $\alpha_1=1$, $\alpha_2=-1$, $\alpha_3=1$, $\alpha_4=-1$, $k=-1$, $m_g(\varepsilon)=0.05$, $g(\varepsilon)=1.1$ and $h(\varepsilon)=1.2$.}\label{skr1A}
    \end{figure}

\begin{figure}[h]
	\centering
	\includegraphics[width=0.8\textwidth]{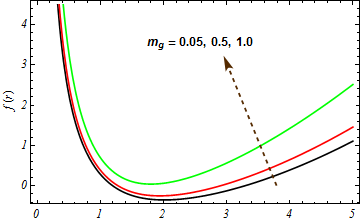}
	\caption{Behaviour of the solution $f(r)$ (Eq. (\ref{17})) for $\Lambda(\varepsilon)=-0.3$, $Q=0.5$, $\beta(\varepsilon)=1.5$, $C=1$, $\mu(\varepsilon)=0.1$, $\alpha_1=1$, $\alpha_2=-1$, $\alpha_3=1$, $\alpha_4=-1$, $k=-1$, $d=4$, $g(\varepsilon)=1.1$ and $h(\varepsilon)=1.2$.}\label{skr1B}
\end{figure}

\begin{figure}[h]
	\centering
	\includegraphics[width=0.8\textwidth]{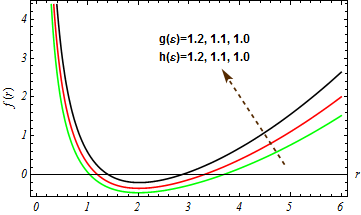}
	\caption{Behaviour of the solution $f(r)$ (Eq. (\ref{17})) for $\Lambda(\varepsilon)=-0.3$, $k=-1$, $Q=0.5$, $\beta(\varepsilon)=1.5$, $C=1$, $\mu(\varepsilon)=0.1$, $\alpha_1=1$, $\alpha_2=-1$, $\alpha_3=1$, $\alpha_4=-1$, $m_g(\varepsilon)=0.05$ and $d=4$.}\label{skr1C}
\end{figure}

\begin{figure}[h]
	\centering
	\includegraphics[width=0.8\textwidth]{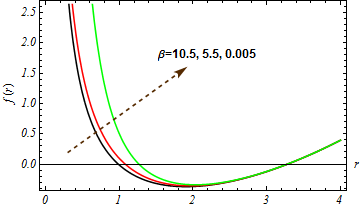}
	\caption{Behaviour of the solution $f(r)$ (Eq. (\ref{17})) for $\Lambda(\varepsilon)=-0.3$, $k=-1$, $Q=0.5$, $d=4$, $C=1$, $\mu(\varepsilon)=0.1$, $\alpha_1=1$, $\alpha_2=-1$, $\alpha_3=1$, $\alpha_4=-1$, $m_g(\varepsilon)=0.05$, $g(\varepsilon)=1.1$ and $h(\varepsilon)=1.2$.}\label{skr1D}
\end{figure}

\begin{figure}[h]
	\centering
	\includegraphics[width=0.8\textwidth]{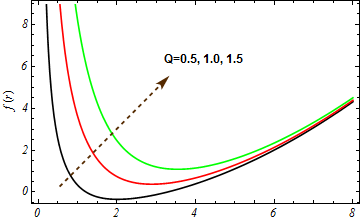}
	\caption{Behaviour of the solution $f(r)$ (Eq. (\ref{17})) for $\Lambda(\varepsilon)=-0.3$, $k=-1$, $\beta(\varepsilon)=1.0$, $d=4$, $C=1$, $\mu(\varepsilon)=0.1$, $\alpha_1=1$, $\alpha_2=-1$, $\alpha_3=1$, $\alpha_4=-1$, $m_g(\varepsilon)=0.05$, $g(\varepsilon)=1.1$ and $h(\varepsilon)=1.2$.}\label{skr1E}
\end{figure}

The Lagrangian density characterizing double-logarithmic electrodynamics \cite{88} is given by
\begin{equation}\begin{split}
L_{m}(\digamma)&=-\frac{1}{2\beta(\varepsilon)}\bigg[\bigg(1-\sqrt{-2\beta(\varepsilon)\digamma}\bigg)\log\bigg(1-\sqrt{-2\beta(\varepsilon)\digamma}\bigg)\\&+\bigg(1+\sqrt{-2\beta(\varepsilon)\digamma}\bigg)\log\bigg(1+\sqrt{-2\beta(\varepsilon)\digamma}\bigg)\bigg], \label{8}\end{split}
\end{equation}
where $\digamma=F_{\mu\nu}F^{\mu\nu}$ defines the Maxwell's invariant in which $F_{\mu\nu}$ represents the Maxwell's field tensor, and $\beta$ is the parameter defines the strength of nonlinearity in electromagnetic field. By varying Eq. (\ref{3}) with respect to the potential $A_{\nu}$, the equations of the electromagnetic field can be constructed as
\begin{equation}
\partial_{\mu}\bigg[\frac{\sqrt{-g}}{\sqrt{-2\beta(\varepsilon)\digamma}}\log\bigg(\frac{1-\sqrt{-2\beta(\varepsilon)\digamma}}{1+\sqrt{-2\beta(\varepsilon)\digamma}}\bigg)F^{\mu\nu}\bigg]=0.
\label{9}
\end{equation}
 The matter-tensor associated with this NLED can be found as
 \begin{equation}\begin{split}
 T^{(m)}_{\mu\nu}&=\big[\big(1-(-2\beta(\varepsilon)\digamma)^{1/2}\big)\log\big(1-(-2\beta(\varepsilon)\digamma)^{1/2}\big)+\big(1+(-2\beta(\varepsilon)\digamma)^{1/2}\big)\\&\times\log\big(1+(-2\beta(\varepsilon)\digamma)^{1/2}\big)\big]\frac{g_{\mu\nu}}{2\beta(\varepsilon)}-\frac{2F_{\mu\lambda}F^{\lambda}_{\nu}}{(-2\beta(\varepsilon)\digamma)^{1/2}}\log\bigg(\frac{1-(-2\beta(\varepsilon)\digamma)^{1/2}}{1+(-2\beta(\varepsilon)\digamma)^{1/2}}\bigg).
 \label{10}\end{split}
 \end{equation}
 The trace of the Eq. (\ref{10}) can be computed as
 \begin{equation}
 T=-\frac{2\digamma}{(-2\beta(\varepsilon)\digamma)^{1/2}}\log\bigg(\frac{1-(-2\beta(\varepsilon)\digamma)^{1/2}}{1+(-2\beta(\varepsilon)\digamma)^{1/2}}\bigg). 
 \label{11}
 \end{equation}
 This shows that the conformal-invariance is violated in this model of NLED. Since, we have considered the line element (\ref{1}) for the energy-dependent spacetime, so it is convenient to introduce the extended fiducial metric as $f_{\mu\nu}=diag(0,0,\frac{C^2h_{ij}}{g^2(\varepsilon)})$, with constant $C>0$ and $h_{ij}$ describes the sector that corresponds to $(d-2)$-dimensional hyper-surface in Eq. (\ref{1}). Taking this form of generalized reference metric, it is easy to compute the polynomials $U_i$ in the form $U_l=\frac{C^l}{r^l}\prod_{a=2}^{l+1}(d-a)$ \cite{13a,144}. One may assume the ansatz for the electromagnetic field tensor in the form \cite{144a,144b,144c,144d}
 \begin{equation}\begin{split}
 F_{\mu\nu}=\frac{2\delta^{\theta_{d-3}}_{[\mu}\delta^{\theta_{d-2}}_{\nu]}q(\varepsilon)^{d-3}}{g^{2d-8}(\varepsilon)r^{d-4}}\times\left\{ \begin{array}{rcl}
 \sin{\theta_{d-3}}\prod_{j=1}^{d-4}\sin^2{\theta_j}, & \mbox{for}
 & k=1 \\1, & \mbox{for} & k=0 \\
 \sin{\theta_{d-3}}\prod_{j=1}^{d-4}\sinh^2{\theta_j}, & \mbox{for} & k=-1
 \end{array}\right.\label{12}
 \end{split}
 \end{equation}
This gives the invariant as
\begin{equation}
\digamma=2q^{2d-6}(\varepsilon)/r^{2d-4}=2Q^2(\varepsilon)/r^{2d-4},\label{13}
\end{equation}
where $q$ and $Q(\varepsilon)$ are the arbitrary constants related to the magnetic charge.
Hence, by putting the line element (\ref{1}), the reference metric $f_{\mu\nu}$, and the matter tensor (\ref{10}) with invariant (\ref{13}) in the gravitational field equations (\ref{6}), one can find the solution in the form
      \begin{equation}\begin{split}
     f(r)&=k-\frac{\mu(\varepsilon)}{r^{d-3}}-\frac{2\Lambda(\varepsilon)r^2}{g^2(\varepsilon)(d-1)(d-2)}+\frac{m^2_g(\varepsilon)}{g^2(\varepsilon)}\bigg(\frac{\alpha_1Cr}{d-2}+\alpha_2C^2+\frac{\alpha_3C^3(d-3)}{r}\\&+\frac{\alpha_4C^4(d-3)(d-4)}{r^2}\bigg)-\frac{4 Q(\varepsilon)(d-3)^{-1}}{g^2(\varepsilon)(d-2)\sqrt{\beta(\varepsilon)}r^{d-4}}\arctan{\bigg(\frac{2\sqrt{\beta(\varepsilon)}Q(\varepsilon)}{r^{d-2}}\bigg)}\\&+\frac{r^2}{\beta(\varepsilon)(d-1)(d-2)g^2(\varepsilon)}\log{\bigg(1+\frac{4\beta(\varepsilon)Q^2(\varepsilon)}{r^{2d-4}}\bigg)}+\frac{8dQ^2(\varepsilon)}{g^2(\varepsilon)(d-3)(d-1)r^{2d-6}}\\&\times F_1\bigg[1,\frac{d-3}{2d-4},\frac{3d-7}{2d-4},-\frac{4\beta(\varepsilon)Q^2(\varepsilon)}{r^{2d-4}}\bigg],\label{17}\end{split}
     \end{equation}
      where $F_1$ is the hypergeometric function while the constant of integration $\mu(\varepsilon)$ is associated with the black hole's mass. The dependence of the metric function on the parameter $\mu(\varepsilon)$ is shown in Fig. \ref{skr1}. The inner and outer horizons can be find out from the equation $f(r)=0$. The values of a coordinate $r$ at which the curve associated with the solution (\ref{17}) crosses the horizontal axis are the location of horizons. The plot of the solution in terms of coordinate $r$ in different dimensional spacetimes is demonstrated in Fig. \ref{skr1A}. It should be noted that the parameter $d$ affects the horizon structure of the energy-dependent magnetized black holes. One can see that for fixed values of other parameters, the solution (\ref{17}) possesses inner and outer horizons in four and five-dimensional spacetimes, however, in higher spacetime dimensional geometries black hole has a single event horizon. Additionally, the solution is plotted for different values of $m_g(\varepsilon)$ in Fig. \ref{skr1B}. It is visualized that the function $f(r)$ describes magnetized black hole with inner and outer horizons when $m_g<m_c$, extremal black hole with single horizon when $m_g(\varepsilon)=m_c$ and naked singularity for $m_g(\varepsilon)>m_c$. Note that $m_c$ is some critical value associated with the mass of graviton. Rainbow functions have also unavoidable effects on the behaviour of the black hole solution (\ref{17}). Fig. \ref{skr1C} shows that the inner (outer) horizon radius increases (decreases) when the rainbow functions are increasing. The solution that has been plotted for $h(\varepsilon)=g(\varepsilon)=1$ corresponds to the black hole of massive gravity. Furthermore, the behaviour of the solution (\ref{17}) in Fig. \ref{skr1D} says that the position of the inner horizon is strongly affected by the nonlinearity of an electromagnetic field, however, the outer horizon radius is the same for any nonzero value of $\beta$. Alternatively, Fig. \ref{skr1E} shows that the positions of the inner and outer horizons are both impacted by differences in charge $Q$.  
   
    The series expansion of the metric function (\ref{17}) about $\beta\rightarrow 0$ is given by
      \begin{equation}\begin{split}
      f(r)&=1-\frac{\mu(\varepsilon)}{r^{d-3}}-\frac{2\Lambda(\varepsilon) r^2}{g^2(\varepsilon)(d-2)(d-1)}+\frac{m^2_g(\varepsilon)}{g^2(\varepsilon)}\big(\frac{\alpha_1Cr}{d-2}+\alpha_2C^2\\&+\frac{\alpha_3C^3(d-3)}{r}+\frac{\alpha_4C^4(d-3)(d-4)}{r^2}\big)+\frac{2(d-3)Q^2(\varepsilon)}{g^2(\varepsilon)(d-2)r^{2d-6}}\\&+O(\beta).\label{18}\end{split}
      \end{equation}
     which represents the anti-de Sitter solution in massive gravity's rainbow coupled to Maxwell's electrodynamics. For the choice $Q=m_g(\varepsilon)=0$, it gives generalized Anti-deSitter Schwarzschild solution in $d$-dimensional gravity's rainbow. However, for $m_g(\varepsilon)=\Lambda(\varepsilon)=0$, it yields the generalized magnetized topological solution with Maxwell's source. Moreover, by taking $h(\varepsilon)=g(\varepsilon)=1$, the solution (\ref{17}) describes the magnetized black holes of massive gravity.
     The Ricci and Kretschmann scalars corresponding to Eq. (\ref{4}) are given by
     \begin{eqnarray}\begin{split}
     R(r)&=g^2(\varepsilon)\bigg[(d^2-5d+6)\frac{(k-f(r))}{r^2}-\frac{d^2f}{dr^2}-\frac{2(d-2)}{r}\frac{df}{dr}\bigg],\label{21}\end{split}
     \end{eqnarray}
     and
     \begin{eqnarray}\begin{split}
     K(r)&=g^4(\varepsilon)\bigg[(2d-4)(d-3)\frac{(k-f(r))^2}{r^4}-\bigg(\frac{d^2f}{dr^2}\bigg)^2+\frac{(2d-4)}{r^2}\bigg(\frac{df}{dr}\bigg)^2\bigg].\label{22}\end{split}
     \end{eqnarray}
     Hence, using the metric function (\ref{17}) in the above expressions, one can show that the above scalars are not regular at $r=0$.

\section{Thermodynamics of Black holes}

The Arnowitt-Deser-Misner (ADM) mass of the asymptotically flat black holes can be determined by considering the large radial distance behaviour of metric functions \cite{145}. However, in the case of anti-de Sitter solutions, the counter-term technique is usually used for the computation of a black hole's finite mass. Note that, both ADM and counter-term techniques yield the finite mass as
\begin{eqnarray}
M=\frac{(d-2)\mu(\varepsilon)V_{d-2}}{16\pi h(\varepsilon)(g(\varepsilon))^{d-1}}.\label{23}
\end{eqnarray}
 From $f(r_h)=0$, it is trouble-free to find the finite mass as
\begin{eqnarray}\begin{split}
M&=\frac{(d-2)V_{d-2}}{16\pi h(\varepsilon)(g(\varepsilon))^{d-3}}\bigg[kr_h^{d-3}-\frac{2\Lambda(\varepsilon)r_h^{d-1}}{g^2(\varepsilon)(d-2)(d-1)}+\frac{m^2_g(\varepsilon)}{(d-2)g^2(\varepsilon)}\bigg(\alpha_1Cr_h^{d-2}\\&+\alpha_2C^2(d-2)r_h^{d-3}+\alpha_3C^3(d-2)(d-3)r_h^{d-4}+\alpha_4C^4(d-2)(d-3)(d-4)r_h^{d-5}\bigg)\\&+\frac{8Q^2(\varepsilon)d(d-2)r_h^{3-d}}{g^2(\varepsilon)(d-3)(d-1)}F_1\bigg[1,\frac{d-3}{2d-4},\frac{3d-7}{2d-4},-\frac{4\beta(\varepsilon)Q^2(\varepsilon)}{r_h^{2d-4}}\bigg]+\frac{r_h^{d-1}}{\beta(\varepsilon)g^2(\varepsilon)(d-1)}\\&\times\log{\bigg(1+\frac{4\beta(\varepsilon)Q^2(\varepsilon)}{r_h^{2d-4}}\bigg)}-\frac{4Q(\varepsilon)r_h}{(d-3)g^2(\varepsilon)\sqrt{\beta(\varepsilon)}}\arctan{\bigg(\frac{2\sqrt{\beta(\varepsilon)}Q(\varepsilon)}{r^{d-2}}\bigg)}\bigg].\label{24}\end{split}
\end{eqnarray}
To inspect the thermodynamical behaviour of the black holes described by Eq. (\ref{17}), it is important to calculate the Hawking temperature. For doing this, one can use the definition of surface gravity according to Hawking \cite{146}, i.e., 
\begin{equation}
T_H=\frac{\kappa_s}{2\pi}, \label{25}
\end{equation}
in which $\kappa_s$ denotes the surface gravity and can be expressed as
\begin{eqnarray}
\label{26}
\begin{split} 
\kappa&=\sqrt{-\frac{1}{2}(\nabla_{\mu}X_{\nu})(\nabla^{\mu}X^{\nu})}, 
\end{split} 
\end{eqnarray}
where $X_{\mu}$ denotes the time-like killing vector field. Hence, one may calculate   
\begin{eqnarray}
\label{27}
\begin{split} 
T_H(r_h)&=\frac{g(\varepsilon)}{4\pi h(\varepsilon)}\bigg[\frac{k(d-3)}{r_h}-\frac{2\Lambda(\varepsilon)r_h}{(d-2)g^2(\varepsilon)}+\frac{m^2_g(\varepsilon)}{g^2(\varepsilon)}\bigg(\frac{\alpha_2C^2}{(d-3)^{-1}r_h}+\frac{\alpha_3C^3(d-3)}{(d-4)^{-1}r_h^2}\\&+\alpha_1C+\frac{\alpha_4C^4(d-3)}{(d-4)^{-1}(d-5)^{-1}r_h^3}\bigg)-\frac{8r_hQ^2(\varepsilon)(d^2-4d+3)^{-1}(d^2-3d-2)}{g^2(\varepsilon)(r_h^{2d-4}+4\beta(\varepsilon) Q^2(\varepsilon))}\\&+\frac{r_h\log{\bigg(1+\frac{4\beta(\varepsilon)Q^2(\varepsilon)}{r_h^{2d-4}}\bigg)}}{(d-2)g^2(\varepsilon)\beta(\varepsilon)}-\frac{4Q(\varepsilon)\arctan{\bigg(\frac{2Q(\varepsilon)\sqrt{\beta(\varepsilon)}}{r_h^{d-2}}\bigg)}}{\sqrt{\beta(\varepsilon)}(d-3)(d-2)g^2(\varepsilon)r_h^{d-3}}\bigg]. 
\end{split} 
\end{eqnarray}
\begin{figure}[h]
	\centering
	\includegraphics[width=0.8\textwidth]{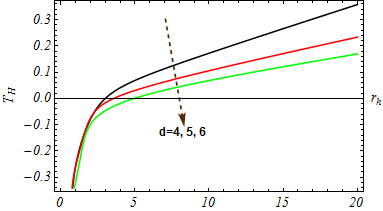}
	\caption{Behaviour of $T_H(r_h)$ (Eq. (\ref{27})) for $\Lambda(\varepsilon)=-0.3$, $Q=1.4$, $\beta(\varepsilon)=1.1$, $C=1$, $\alpha_1=1$, $\alpha_2=-1$, $\alpha_3=1$, $\alpha_4=-1$, $m_g(\varepsilon)=0.05$, $g(\varepsilon)=1.1$, $k=-1$ and $h(\varepsilon)=1.2$.}\label{skr3}
\end{figure} 
\begin{figure}[h]
	\centering
	\includegraphics[width=0.8\textwidth]{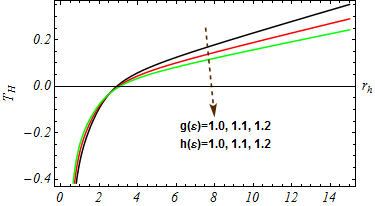}
	\caption{Behaviour of $T_H(r_h)$ (Eq. (\ref{27})) for $\Lambda(\varepsilon)=-0.3$, $Q=1.4$, $\beta(\varepsilon)=1.1$, $C=1$, $\alpha_1=1$, $\alpha_2=-1$, $\alpha_3=1$, $\alpha_4=-1$, $m_g(\varepsilon)=0.05$, $k=-1$ and $d=4$.}\label{skr3A}
\end{figure}
\begin{figure}[h]
	\centering
	\includegraphics[width=0.8\textwidth]{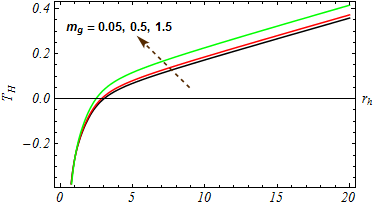}
	\caption{Behaviour of $T_H(r_h)$ (Eq. (\ref{27})) for $\Lambda(\varepsilon)=-0.3$, $Q=1.4$, $\beta(\varepsilon)=1.1$, $C=1$, $\alpha_1=1$, $\alpha_2=-1$, $\alpha_3=1$, $\alpha_4=-1$, $d=4$, $g(\varepsilon)=1.1$, $k=-1$ and $h(\varepsilon)=1.2$.}\label{skr3B}
\end{figure} \begin{figure}[h]
\centering
\includegraphics[width=0.8\textwidth]{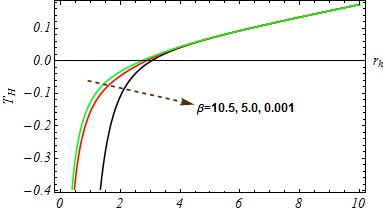}
\caption{Behaviour of $T_H(r_h)$ (Eq. (\ref{27})) for $\Lambda(\varepsilon)=-0.3$, $Q=1.5$, $d=4$, $C=1$, $\alpha_1=1$, $\alpha_2=-1$, $\alpha_3=1$, $\alpha_4=-1$, $m_g(\varepsilon)=0.05$, $g(\varepsilon)=1.1$, $k=-1$ and $h(\varepsilon)=1.2$.}\label{skr3C}
\end{figure}
\begin{figure}[h]
	\centering
	\includegraphics[width=0.8\textwidth]{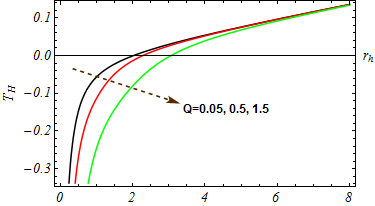}
	\caption{Behaviour of $T_H(r_h)$ (Eq. (\ref{27})) for $\Lambda(\varepsilon)=-0.3$, $\beta(\varepsilon)=1.5$, $d=4$, $C=1$, $\alpha_1=1$, $\alpha_2=-1$, $\alpha_3=1$, $\alpha_4=-1$, $m_g(\varepsilon)=0.05$, $g(\varepsilon)=1.1$, $k=-1$ and $h(\varepsilon)=1.2$.}\label{skr3D}
\end{figure}  
The behaviour of Hawking temperature for various values of dimensionality parameter and rainbow functions can be examined in Figs. \ref{skr3} and \ref{skr3A}. The type 1 phase transition is indicated by the point at which temperature changes from a negative to a positive sign. Similarly, its positivity expresses the physicality of gravitating object. One can conclude that four-dimensional black holes are more physical objects than the black holes in higher dimensions. For a fixed value of $d$, it is shown that the temperature falls as the values of the rainbow functions grow. Additionally, the curve corresponding to $g(\varepsilon)=h(\varepsilon)=1$ in Fig. \ref{skr3A} depicts the Hawking temperature of the magnetized black hole in massive gravity. Fig. \ref{skr3B} shows that the temperature is increasing with the increasing of parameter $m_g(\varepsilon)$. Similarly, the effects of the nonlinearity and magnetic charge parameters on the physicality of black holes are investigated in Figs. \ref{skr3C} and \ref{skr3D}. It is shown that the outer horizon radius associated with the phase transition of type 1 becomes smaller when the nonlinearity parameter is increasing, On the other hand, when a magnetic charge increases, the value of this critical horizon corresponding to the physical object also increases. Furthermore, the case $Q(\varepsilon)=0$ in Eq. (\ref{27}) describes the Hawking temperature of the neutral black hole in massive gravity's rainbow and for $m_g(\varepsilon)=0$, it yields the temperature of the black hole in gravity's rainbow. 
 The entropy \cite{147,148} can be computed from the area law as
 \begin{eqnarray}\begin{split}
 S=\frac{V_{d-2}r_h^{d-2}}{4(g(\varepsilon))^{d-2}}.\label{27a}\end{split}
 \end{eqnarray}
  Using eq. (\ref{27a}), it is also possible to express the finite mass (\ref{24}) as a function of entropy $S$ as
  \begin{eqnarray}\begin{split}
  M(S_1)&=\frac{(d-2)V_{d-2}}{16\pi h(\varepsilon)(g(\varepsilon))^{d-3}}\bigg[k g(\varepsilon)^{d-3}S_1^{\frac{d-3}{d-2}}-\frac{2\Lambda(\varepsilon)S_1^{\frac{d-1}{d-2}}}{(d-2)(d-1)g^{2}(\varepsilon)}+\frac{m^2_g(\varepsilon)}{(d-2)g^2(\varepsilon)}\\&\bigg(\alpha_1C (g(\varepsilon))^{d-2}S_1+\alpha_2C^2(d-2)(g(\varepsilon))^{d-3}S_1^{\frac{d-3}{d-2}}+\alpha_3C^3(d-2)(d-3)(g(\varepsilon))^{d-4}\\&\times S_1^{\frac{d-4}{d-2}}+\alpha_4C^4(d-2)(d-3)(d-4)(g(\varepsilon))^{d-5}S_1^{\frac{d-5}{d-2}}\bigg)+\frac{8Q^2d(d-2)Q^2(\varepsilon)}{(g(\varepsilon))^{d-1}(d-3)}\\&\times\frac{S_1^{\frac{3-d}{d-2}}}{(d-1)} F_1\bigg[1,\frac{d-3}{2d-4},\frac{3d-7}{2d-4},-\frac{4\beta(\varepsilon)Q^2(\varepsilon)}{(g(\varepsilon))^{2d-4}S_1^2}\bigg]+\frac{(g(\varepsilon))^{d-3}S_1^{\frac{d-1}{d-2}}}{\beta(\varepsilon)(d-1)(d-2)}\\&\times\log{\bigg(1+\frac{4\beta(\varepsilon)Q^2(\varepsilon)}{(g(\varepsilon))^{2d-4}S_1^2}\bigg)}-\frac{4Q(\varepsilon)S_1^{\frac{1}{d-2}}}{g(\varepsilon)(d-3)\sqrt{\beta(\varepsilon)}}\arctan{\bigg(\frac{2\sqrt{\beta(\varepsilon)}Q(\varepsilon)}{S_1g^{d-2}(\varepsilon)}\bigg)}\bigg],\label{28}\end{split}
  \end{eqnarray}
where we have used $S_1=4S/V_{d-2}$ for simplicity. The Smarr's relation can be obtained by using Euler's homogeneous function theorem which states that when
$f(\xi^ix,\xi^jy,\xi^kz)=\xi^lf(x,y,z)$ such that $\xi$ is a constant and $i,j,k$ are the integers, then 
\begin{equation}
lf(x,y,z)=ix\frac{\partial f}{\partial x}+jy\frac{\partial f}{\partial y}+kz\frac{\partial f}{\partial z}. \label{29}
\end{equation}
Using Eq. (\ref{28}) and the rescaling of variables involved in the finite mass, i.e.,
\begin{eqnarray}\begin{split}
S_1&\rightarrow \xi^iS_1, Q(\varepsilon)\rightarrow	\xi^j Q(\varepsilon), \beta(\varepsilon)\rightarrow \xi^k \beta(\varepsilon)\\&\Lambda(\varepsilon)\rightarrow \xi^n\Lambda(\varepsilon), \alpha_i \rightarrow \xi^{r_u}\alpha_i, \label{30}\end{split}
	\end{eqnarray}
where $u=1,2,3,4$, one can see that the Euler's homogeneous function theorem holds if the integer powers satisfy the following equalities
 \begin{eqnarray}\begin{split}
(d-3)i&=(d-2)l, j=l, (d-3)k=2l, (d-3)r_1=-l, r_2=0,\\& (d-3)r_3=l, (d-3)r_4=2l.\label{31}\end{split}
 \end{eqnarray}
Hence, from the above choices of integer powers, one can construct the Smarr's formula as
\begin{equation}\begin{split}
(d-3)M[S,Q(\varepsilon),\beta(\varepsilon),\Lambda(\varepsilon),m_g(\varepsilon)]& =(d-2)ST_H+(d-3)Q(\varepsilon)\Phi_Q+2\beta(\varepsilon)W_{\beta}\\&+2PV(\varepsilon)-\alpha_1B_1+\alpha_3B_3+2\alpha_4B_4,
,\label{32}\end{split}\end{equation}
where
\begin{eqnarray}\begin{split}
T_H(S)&=\frac{\partial M}{\partial S}=\frac{g(\varepsilon)}{4\pi h(\varepsilon)(d-2)}\bigg[\frac{k(d-3)V_{d-2}^{\frac{1}{d-2}}}{g(\varepsilon)4^{\frac{1}{d-2}}S^{\frac{1}{d-2}}}-\frac{4^{\frac{1}{d-2}}\Lambda(\varepsilon)S^{\frac{1}{d-2}}}{2^{-1}g(\varepsilon)(d-2)V^{\frac{1}{d-2}}}\\&+\frac{m^2_g(\varepsilon)}{g^2(\varepsilon)(d-2)}\bigg(\alpha_1C+\frac{(d-3)V_{d-2}^{\frac{1}{d-2}}\alpha_2C^2}{4^{\frac{1}{d-2}}g(\varepsilon)S^{\frac{1}{d-2}}}+\frac{(d-3)(d-4)V_{d-2}^{\frac{2}{d-2}}\alpha_3C^3}{g^2(\varepsilon)4^{\frac{2}{d-2}}S^{\frac{2}{d-2}}}\\&+\frac{\alpha_4C^4(d-3)(d^2-9d+20)V_{d-2}^{\frac{3}{d-2}}}{4^{\frac{3}{d-2}}g^3(\varepsilon)S^{\frac{3}{d-2}}}\bigg)-\frac{4^{\frac{d-1}{d-2}}Q^2(\varepsilon)(d-2)S^{\frac{1}{d-2}}}{(d-1)V_{d-2}^{\frac{5-2d}{d-2}}g(\varepsilon)}\\&\times\frac{1}{(16g^{2d-4}S^2+4\beta Q^2V^2_{d-2})}+\frac{2Q^2(\varepsilon)(d-5)V_{d-2}^{\frac{d-3}{d-2}}}{4^{\frac{d-3}{d-2}}(d-2)g^{d-1}(\varepsilon)\sqrt{\beta(\varepsilon)}S^{\frac{d-3}{d-2}}}\\&\times\arctan{\bigg(\frac{V_{d-2}Q(\varepsilon)\sqrt{\beta(\varepsilon)}}{2g^{d-2}(\varepsilon)S}\bigg)}+\frac{4^{\frac{1}{d-2}}S^{\frac{1}{d-2}}}{(d-2)\beta(\varepsilon)g(\varepsilon)V_{d-2}^{\frac{1}{d-2}}}\\&\log{\bigg(1+\frac{\beta Q^2V_{d-2}^2}{4g^{2d-4}(\varepsilon)S^2}\bigg)}\bigg],\label{33}\end{split}
\end{eqnarray}
defines the Hawking temperature in terms of entropy $S$. The quantity $\Phi_Q(S)$ refers to the magnetic potential given by
\begin{eqnarray}\begin{split}
\Phi_Q(S)&=\frac{\partial M}{\partial Q}=\frac{V_{d-2}^{\frac{d-1}{d-2}}S^{\frac{1}{d-2}}(4g^{2d-4}(\varepsilon)+\beta(\varepsilon)Q^(\varepsilon)V^2_{d-2})^{-1}}{16\pi(d-1)(d-3)h(\varepsilon)g^{d-4}(\varepsilon)}\bigg[\frac{g^d(\varepsilon)(d-3)}{\sqrt{\beta(\varepsilon)}}\\&\times\bigg(4^{\frac{d-1}{d-2}}g^{d+2}(\varepsilon)Q(\varepsilon)V_{d-2}^{\frac{d-4}{d-2}}\sqrt{\beta(\varepsilon)}S^{\frac{d}{d-2}}\bigg(2V_{d-2}^{\frac{d}{d-2}}-(d-1)\bigg)-\frac{(d-1)2^{\frac{d}{d-2}}}{V_{d-2}^{\frac{2}{d-2}}}\\&\times g^4(\varepsilon)S^{\frac{d}{d-2}}\bigg(4g^{2d-4}(\varepsilon)S^2+\beta Q^2V^2_{d-2}\bigg)\cot^{-1}{\bigg(\frac{2g^{d-2}(\varepsilon)S}{Q(\varepsilon)V_{d-2}\sqrt{\beta(\varepsilon)}}\bigg)}\bigg)\\&+\frac{(d-2)Q(\varepsilon)g^2(\varepsilon)V_{d-2}^{\frac{d-4}{d-2}}}{S^{\frac{d-4}{d-2}}}\bigg(4^{\frac{d-1}{d-2}}(d-3)g^{2d}(\varepsilon)S^2-\bigg(2^{\frac{2}{d-2}}(d-3)\\&-(d-2)2^{\frac{d}{d-2}}\bigg)\bigg(4g^{2d-4}(\varepsilon)+Q^2(\varepsilon)\beta(\varepsilon)V^2_{d-2}\bigg)g^4(\varepsilon)F_1\bigg[1,\frac{d-3}{2d-4},\\&\frac{3d-7}{2d-4},\frac{-g^{2d-4}(\varepsilon)\beta(\varepsilon)V^2_{d-2}Q^2(\varepsilon)}{4S^2}\bigg]\bigg)\bigg].\label{34}\end{split}
\end{eqnarray}
Similarly, the conjugate quantity $W_{\beta}$ associated with the nonlinearity parameter can be found as
\begin{eqnarray}\begin{split}
	W_{\beta}(S)&=\frac{\partial M}{\partial \beta}=\frac{V_{d-2}(d-2)}{16\pi h(\varepsilon)g^{d-1}(\varepsilon)}\bigg[\frac{4^{\frac{1}{d-2}}Q^2(\varepsilon)g^{1-d}(\varepsilon)S^{\frac{d-1}{d-2}}}{(d-1)(d-2)\beta(\varepsilon)V_{d-2}^{\frac{1}{d-2}}}\\&\times\bigg(S^2+\frac{\beta(\varepsilon)Q^2(\varepsilon)V_{d-2}^2}{4g^{2d-4}(\varepsilon)}\bigg)^{-1}+\frac{2^{\frac{d}{d-2}}Q(\varepsilon)(d-3)S^{\frac{1}{d-2}}}{g(\varepsilon)V_{d-2}^{\frac{1}{d-2}}\beta^{\frac{3}{2}}(\varepsilon)}\\&\times \arctan{\bigg(\frac{Q(\varepsilon)V_{d-2}\sqrt{\beta(\varepsilon)}}{2g^{d-2}(\varepsilon)S}\bigg)}-\frac{4^{\frac{d-1}{d-2}}g^{d-3}(\varepsilon)S^{\frac{d-1}{d-2}}}{(d-1)(d-2)V_{d-2}^{\frac{d-1}{d-2}}\beta^2(\varepsilon)}\\&\times\log{\bigg(1+\frac{\beta(\varepsilon)Q^2(\varepsilon)V^2_{d-2}}{4g^{2d-4}(\varepsilon)S^2}\bigg)}-\frac{Q^2(\varepsilon)V^{\frac{1}{d-3}}}{2^{\frac{d}{d-2}}\beta(\varepsilon)(d-2)g^{d-1}(\varepsilon)}\\&\times\frac{4S^{\frac{d-1}{d-2}}g^{2d-4}(\varepsilon)}{4g^{2d-4}(\varepsilon)S^2+\beta(\varepsilon)Q^2(\varepsilon)V_{d-2}^2}+\frac{Q^2(\varepsilon)}{2^{\frac{d-4}{d-2}}(d-1)\beta(\varepsilon)}\\&\times\frac{S^{\frac{d-3}{d-2}}}{g^{d-1}(\varepsilon)}\bigg(\bigg(1+\frac{\beta(\varepsilon)Q^2V^2_{d-2}}{4g^{2d-4}(\varepsilon)S^2}\bigg)^{-1}-F_1\bigg[1,\frac{d-3}{2d-4},\frac{3d-7}{2d-4},\\&\frac{-g^{2d-4}(\varepsilon)\beta(\varepsilon)V^2_{d-2}Q^2(\varepsilon)}{4S^2}\bigg]\bigg)\bigg].\label{35}\end{split}
\end{eqnarray}
The quantity $V$ which designates the thermodynamic volume conjugate to pressure $P=-\Lambda(\varepsilon)$ can be computed as
\begin{eqnarray}\begin{split}
V(S)&=\frac{\partial M}{\partial P}=\frac{4^{\frac{d-1}{d-2}}S^{\frac{d-1}{d-2}}}{8\pi h(\varepsilon)g^{d-1}(\varepsilon)(d-1)V_{d-2}^{\frac{1}{d-2}}}.\label{36}\end{split}
\end{eqnarray}
Finally, the conjugate quantities $B_i$'s associated with $\alpha_i$'s are calculated as 
 \begin{eqnarray}\begin{split}
 B_{1}(S)&=\frac{\partial M}{\partial \alpha_1}=\frac{m^2_g(\varepsilon)CS}{4\pi h(\varepsilon)g(\varepsilon)},\label{37}\end{split}
 \end{eqnarray}
 \begin{eqnarray}\begin{split}
 B_{2}(S)&=\frac{\partial M}{\partial \alpha_2}=\frac{4^{\frac{d-3}{d-2}}m^2_g(\varepsilon)(d-2)C^2S^{\frac{d-3}{d-2}}}{16\pi h(\varepsilon)g^2(\varepsilon)V_{d-2}^{\frac{-1}{d-2}}},\label{38}\end{split}
 \end{eqnarray}
 \begin{eqnarray}\begin{split}
 B_{3}(S)&=\frac{\partial M}{\partial \alpha_3}=\frac{4^{\frac{d-4}{d-2}}m^2_g(\varepsilon)(d-2)(d-3)C^3S^{\frac{d-4}{d-2}}}{16\pi h(\varepsilon)g^3(\varepsilon)V_{d-2}^{\frac{-2}{d-2}}},\label{39}\end{split}
 \end{eqnarray}
 and
 \begin{eqnarray}\begin{split}
 B_{4}(S)&=\frac{\partial M}{\partial \alpha_4}=\frac{4^{\frac{d-5}{d-2}}C^4m^2_g(\varepsilon)(d-2)(3-d)(4-d)S^{\frac{d-5}{d-2}}}{16\pi h(\varepsilon)g^4(\varepsilon)V_{d-2}^{\frac{-3}{d-2}}}.\label{40}\end{split}
 \end{eqnarray}
  Using the above expression of mass, one can construct the first law of thermodynamics as
 \begin{eqnarray}\begin{split}
 dM=T_HdS+\Phi_Q dQ+W_{\beta}d\beta+Vd\Lambda+\sum_{i=1}^{4}B_id\alpha_i,\label{40a}\end{split}
 \end{eqnarray}
 The heat capacity can be defined as
\begin{eqnarray}\begin{split}
C_Q=T_H(r_h)\frac{\partial S}{\partial r_h}\bigg(\frac{\partial T_H}{\partial r_h}\bigg)^{-1}|_{Q},\label{41}\end{split}
\end{eqnarray}
Hence, by using Eqs. (\ref{27}) and (\ref{27a}) in Eq. (\ref{41}), one obtains
\begin{eqnarray}\begin{split}
C_Q&=\frac{(d-2)V_{d-2}r_h^{d-3}\bigg(\frac{k(d-3)}{r_h}-\frac{2\Lambda(\varepsilon)r_h}{g^2(\varepsilon)(d-2)}+\Delta_1(r_h)+\Delta_2(r_h)\bigg)}{4g^{d-2}(\varepsilon)\bigg(\Theta_(r_h)+\Theta_2(r_h)-\frac{k(d-3)}{r_h^2}-\frac{2\Lambda(\varepsilon)}{g^2(\varepsilon)(d-2)}\bigg)},\label{43}\end{split}\end{eqnarray}
where
\begin{eqnarray}\begin{split}
\Delta_1(r_h)&=\frac{m^2_g(\varepsilon)}{g^2(\varepsilon)}\bigg[\frac{\alpha_2C^2(d-3)}{r_h}+\frac{\alpha_3C^3(d^2-7d+12)}{r_h^2}+\alpha_1C\\&+\frac{\alpha_4C^4(d-3)(d^2-9d+20)}{r_h^3}\bigg],\label{44}\end{split}\end{eqnarray}
\begin{eqnarray}\begin{split}
\Delta_2(r_h)&=-\frac{8r_hQ^2(\varepsilon)(d^2-3d-2)}{g^2(\varepsilon)(d^2-4d+3)(r_h^{2d-4}+4\beta(\varepsilon) Q^2(\varepsilon))}\\&+\frac{r_h\log{\bigg(1+\frac{4\beta(\varepsilon)Q^2(\varepsilon)}{r_h^{2d-4}}\bigg)}}{(d-2)g^2(\varepsilon)\beta(\varepsilon)}-\frac{4Q(\varepsilon)\arctan{\bigg(\frac{2Q(\varepsilon)\sqrt{\beta(\varepsilon)}}{r_h^{d-2}}\bigg)}}{\sqrt{\beta(\varepsilon)}(d-3)(d-2)g^2(\varepsilon)r_h^{d-3}},\label{45}\end{split}\end{eqnarray}
\begin{eqnarray}\begin{split}
\Theta_1(r_h)&=-\frac{m^2_g(\varepsilon)}{g^2(\varepsilon)}\bigg(\frac{(d-3)\alpha_2C^2}{r_h^2}+\frac{2(d-3)(d-4)\alpha_3C^3}{r_h^3}\\&+\frac{3(d-3)(d-4)(d-5)\alpha_4C^4}{r_h^4}\bigg),\label{46}\end{split}\end{eqnarray}
and
\begin{eqnarray}\begin{split}
\Theta_2(r_h)&=\frac{16(d-2)(d^2-3d-2)Q^2(\varepsilon)r_h^{2d-4}}{(d-1)(d-3)g^2(\varepsilon)(r_h^{2d-4}+4Q^2(\varepsilon)\beta(\varepsilon))^2}+\frac{4Q(\varepsilon)}{(d-2)g^2(\varepsilon)\sqrt{\beta}r_h^{d-2}}\\&\times\arctan{\bigg(\frac{2Q(\varepsilon)\sqrt{\beta}}{r_h^{d-2}}\bigg)}+\frac{1}{(d-2)\beta(\varepsilon)g^2(\varepsilon)}\log{\bigg(\frac{r_h^{2d-4}+4\beta(\varepsilon)Q^2(\varepsilon)}{r_h^{2d-4}}\bigg)}\\&-\frac{16Q^2(\varepsilon)(d^2-4d+2)}{(d-1)(d-3)g^2(\varepsilon)(r_h^{2d-4}+4Q^2(\varepsilon)\beta(\varepsilon))}.\label{47}\end{split}\end{eqnarray}
\begin{figure}[h]
	\centering
	\includegraphics[width=0.8\textwidth]{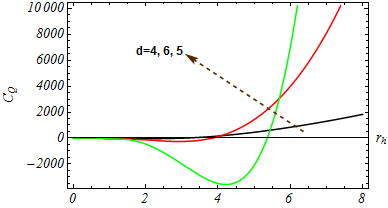}
	\caption{Dependence of $C_Q(r_h)$ (Eq. (\ref{43})) on the dimensionality parameter $d$. The other parameters are selected as $\Lambda(\varepsilon)=-0.3$, $Q=1.4$, $\beta(\varepsilon)=1.1$, $C=1$, $\alpha_1=1$, $\alpha_2=-1$, $\alpha_3=1$, $\alpha_4=-1$, $m_g(\varepsilon)=0.05$, $g(\varepsilon)=1.2$, $k=-1$ and $h(\varepsilon)=1.1$.}\label{skr4}
\end{figure} 
\begin{figure}[h]
	\centering
	\includegraphics[width=0.8\textwidth]{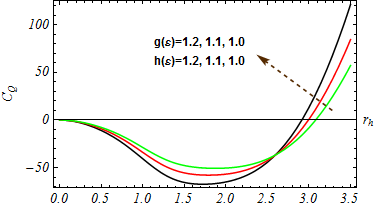}
	\caption{Dependence of $C_Q(r_h)$ (Eq. (\ref{43})) on the rainbow functions. The other parameters are selected as $\Lambda(\varepsilon)=-0.3$, $Q=1.4$, $\beta(\varepsilon)=1.1$, $C=1$, $\alpha_1=1$, $\alpha_2=-1$, $\alpha_3=1$, $\alpha_4=-1$, $m_g(\varepsilon)=0.05$, $k=-1$ and $d=4$.}\label{skr4A}
\end{figure}
\begin{figure}[h]
	\centering
	\includegraphics[width=0.8\textwidth]{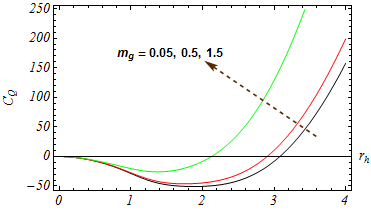}
	\caption{Dependence of $C_Q(r_h)$ (Eq. (\ref{43})) on the parameter $m_g$. The other parameters are selected as $\Lambda(\varepsilon)=-0.3$, $Q=1.4$, $\beta(\varepsilon)=1.1$, $C=1$, $\alpha_1=1$, $\alpha_2=-1$, $\alpha_3=1$, $\alpha_4=-1$, $d=4$, $g(\varepsilon)=1.2$, $k=-1$ and $h(\varepsilon)=1.1$.}\label{skr4B}
\end{figure}
 \begin{figure}[h]
 	\centering
 	\includegraphics[width=0.8\textwidth]{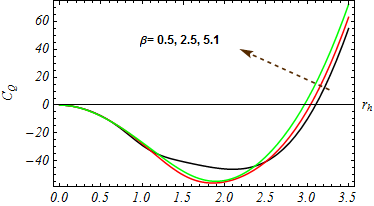}
 	\caption{Dependence of $C_Q(r_h)$ (Eq. (\ref{43})) on the nonlinearity of the electromagnetic field. The other parameters are selected as $\Lambda(\varepsilon)=-0.3$, $Q=1.4$, $d=4$, $C=1$, $\alpha_1=1$, $\alpha_2=-1$, $\alpha_3=1$, $\alpha_4=-1$, $m_g(\varepsilon)=0.05$, $g(\varepsilon)=1.2$, $k=-1$ and $h(\varepsilon)=1.1$.}\label{skr4C}
 \end{figure}
\begin{figure}[h]
	\centering
	\includegraphics[width=0.8\textwidth]{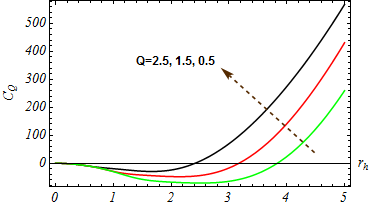}
	\caption{Dependence of $C_Q(r_h)$ (Eq. (\ref{43})) on the nonlinearity of the electromagnetic field. The other parameters are selected as $\Lambda(\varepsilon)=-0.3$, $\beta(\varepsilon)=0.5$, $d=4$, $C=1$, $\alpha_1=1$, $\alpha_2=-1$, $\alpha_3=1$, $\alpha_4=-1$, $m_g(\varepsilon)=0.05$, $g(\varepsilon)=1.2$, $k=-1$ and $h(\varepsilon)=1.1$.}\label{skr4D}
\end{figure}

Figs. \ref{skr4} and \ref{skr4A} describes the behaviour of the heat capacity for different values of the dimensionality parameter $d$ and rainbow functions, respectively. The region in which it is nonnegative illustrates the local stability of the resulting magnetized black holes described by Eq. (\ref{17}). The point at which it vanishes indicates the existence of the type 1 phase transition, however, the point at which it fails to be analytic corresponds to the phase transition of type 2. One can detect that the size of the stable black hole increases in higher dimensional spacetimes. It is also observed that the smaller values of the rainbow functions lead towards a larger region of $r_h$ associated with local stability in a fixed spacetime dimension. On the other hand, the critical horizon radius associated with type 1 phase transitions becomes larger when the rainbow functions are increasing above unity. The dependence of local thermodynamic stability on the parameter $m_g(\varepsilon)$ is demonstrated in Fig. \ref{skr4B}. It is shown that the specific heat is positive in a larger region of $r_h$ when the greater value of $m_g(\varepsilon)$ is chosen. Hence, the magnetized black holes in the presence of massive gravitons are more locally stable. Furthermore, the heat capacity for different values of the nonlinearity parameter $\beta(\varepsilon)$ and magnetic charge $Q$ is plotted in Figs. \ref{skr4C} and \ref{skr4D}, respectively. One can see that the horizon radius $r_h$ associated with the smallest stable black hole decreases when the parameter describing the nonlinearity of an electromagnetic field increases. This shows that the nonlinear magnetically charged black holes are more locally stable than the black holes sourced by the Maxwell field. However, if the other parameters are fixed, the objects with a smaller magnitude of charge $Q$ are more stable. It should be noted that the case $Q=0$ in (\ref{43}) corresponds to the heat capacity of uncharged black holes in the massive gravity's rainbow, whereas the substitution of $m_g(\varepsilon)=0$ in this equation results in the heat capacity of black holes in doubly general relativity or gravity's rainbow. Furthermore, the curve based on the values $g(\varepsilon)=h(\varepsilon)=1$ in Fig. \ref{skr4A} describes the behaviour of the specific heat of the magnetized black hole in massive gravity.

 The global thermodynamic stability can be checked from the behaviour of Gibbs free energy in terms of $r_h$ \cite{149}. This quantity is given by
\begin{equation}
G=M-T_HS, \label{48}
\end{equation}
which by the substitution of mass (\ref{24}), temperature (\ref{27}) and entropy (\ref{27a}) gives the expression
\begin{eqnarray}\begin{split}
G&=\frac{(d-2)V_{d-2}}{16\pi h(\varepsilon)(g(\varepsilon))^{d-3}}\bigg(kr_h^{d-3}-\frac{2\Lambda(\varepsilon)r_h^{d-1}}{g^2(\varepsilon)(d-2)(d-1)}+\Xi_1(r_h)+\Xi_2(r_h)\bigg)\\&-\frac{V_{d-2}r_h^{d-2}}{16\pi h(\varepsilon)g(\varepsilon)^{d-3}}\bigg(\frac{k(d-3)}{r_h}-\frac{2\Lambda(\varepsilon)r_h}{g^2(\varepsilon)(d-2)}+\Delta_1(r_h)+\Delta_2(r_h)\bigg),\label{49}\end{split}
\end{eqnarray}
where 
\begin{eqnarray}\begin{split}
\Xi_1(r_h)&=\frac{m^2_g(\varepsilon)}{(d-2)g^2(\varepsilon)}\bigg(\alpha_1Cr_h^{d-2}+\alpha_2C^2(d-2)r_h^{d-3}+\alpha_3C^3(d-2)(d-3)r_h^{d-4}\\&+\alpha_4C^4(d-2)(d-3)(d-4)r_h^{d-5}\bigg),\label{50}\end{split}\end{eqnarray}
and
\begin{eqnarray}\begin{split}
\Xi_2(r_h)&=\frac{8Q^2(\varepsilon)d(d-2)r_h^{3-d}}{g^2(\varepsilon)(d-3)(d-1)}F_1\bigg[1,\frac{d-3}{2d-4},\frac{3d-7}{2d-4},-\frac{4\beta(\varepsilon)Q^2(\varepsilon)}{r_h^{2d-4}}\bigg]+\frac{(d-1)^{-1}r_h^{d-1}}{\beta(\varepsilon)g^2(\varepsilon)}\\&\times\log{\bigg(1+\frac{4\beta(\varepsilon)Q^2(\varepsilon)}{r_h^{2d-4}}\bigg)}-\frac{4Q(\varepsilon)r_h}{(d-3)g^2(\varepsilon)\sqrt{\beta(\varepsilon)}}\arctan{\bigg(\frac{2\sqrt{\beta(\varepsilon)}Q(\varepsilon)}{r^{d-2}}\bigg)}.\label{51}\end{split}\end{eqnarray}
Note that, the values of $\Delta_1(r_h)$ and $\Delta_2(r_h)$ are given in Eqs. (\ref{45}) and (\ref{46}), respectively.

\begin{figure}[h]
	\centering
	\includegraphics[width=0.8\textwidth]{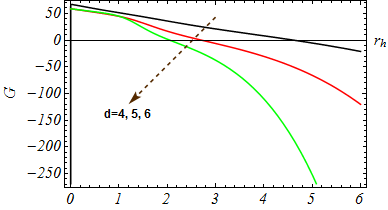}
	\caption{Behaviour of $G(r_h)$ (Eq. (\ref{49})) for $\Lambda(\varepsilon)=-0.3$, $Q=1.4$, $\beta(\varepsilon)=1.1$, $C=1$, $\alpha_1=1$, $\alpha_2=-1$, $\alpha_3=1$, $\alpha_4=-1$, $m_g(\varepsilon)=0.05$, $k=-1$, $g(\varepsilon)=1.1$ and $h(\varepsilon)=1.2$.}\label{skr5}
\end{figure}
\begin{figure}[h]
	\centering
	\includegraphics[width=0.8\textwidth]{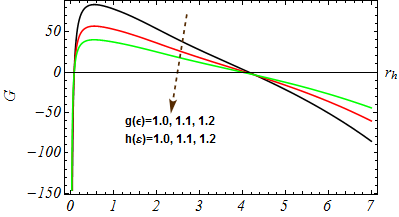}
	\caption{Behaviour of $G(r_h)$ (Eq. (\ref{49})) for $\Lambda(\varepsilon)=-0.3$, $Q=1.4$, $\beta(\varepsilon)=1.1$, $C=1$, $\alpha_1=1$, $\alpha_2=-1$, $\alpha_3=1$, $\alpha_4=-1$, $m_g(\varepsilon)=1.5$, $k=-1$ and $d=5$.}\label{skr5A}
\end{figure}
\begin{figure}[h]
	\centering
	\includegraphics[width=0.8\textwidth]{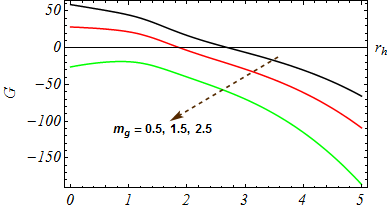}
	\caption{Behaviour of $G(r_h)$ (Eq. (\ref{49})) for $\Lambda(\varepsilon)=-0.3$, $Q=1.4$, $\beta(\varepsilon)=1.1$, $C=1$, $\alpha_1=1$, $\alpha_2=-1$, $\alpha_3=1$, $\alpha_4=-1$, $k=-1$, $d=5$, $g(\varepsilon)=1.1$ and $h(\varepsilon)=1.2$.}\label{skr5B}
\end{figure}
\begin{figure}[h]
	\centering
	\includegraphics[width=0.8\textwidth]{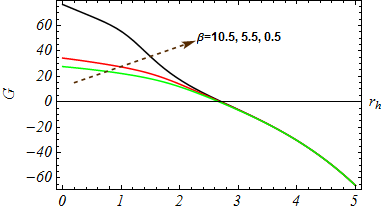}
	\caption{Behaviour of $G(r_h)$ (Eq. (\ref{49})) for $\Lambda(\varepsilon)=-0.3$, $Q=1.4$, $d=5$, $C=1$, $\alpha_1=1$, $\alpha_2=-1$, $\alpha_3=1$, $\alpha_4=-1$, $m_g(\varepsilon)=0.05$, $k=-1$, $g(\varepsilon)=1.1$ and $h(\varepsilon)=1.2$.}\label{skr5C}
\end{figure}
\begin{figure}[h]
	\centering
	\includegraphics[width=0.8\textwidth]{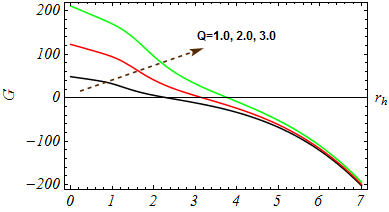}
	\caption{Behaviour of $G(r_h)$ (Eq. (\ref{49})) for $\Lambda(\varepsilon)=-0.3$, $d=5$, $\beta(\varepsilon)=0.5$, $C=1$, $\alpha_1=1$, $\alpha_2=-1$, $\alpha_3=1$, $\alpha_4=-1$, $m_g(\varepsilon)=0.05$, $k=-1$, $g(\varepsilon)=1.1$ and $h(\varepsilon)=1.2$.}\label{skr5D}
\end{figure}

 The behaviour of Gibbs free energy associated with different dimensional black holes can be noticed in Fig. \ref{skr5}. Those values of horizon radius $r_h$ that correspond to the negative values of this quantity show that the black hole of such horizon radii is globally stable. The regions where the objects described by (\ref{17}) are enjoying global thermodynamic stability are demonstrated in Fig. \ref{skr5}. In addition to this, one can investigate the effects of rainbow functions on the behaviour of Gibbs free energy in Fig. \ref{skr5A}. There exists three intervals of $r_h$, say, $(0, r_1)$, $(r_1,r_2)$ and $(r_2,\infty)$ where these effects can be visualized. In the interval $(0, r_1)$, Gibbs energy is negative and increasing until $r_h$ reaches $r_1$. Thus, the black hole with $r_h$ belongs to $(0, r_1)$ is globally stable. In the second interval $(r_1,r_2)$, it is positive and so the black hole with horizon radius $r_h$ within this interval is globally unstable. However, in the third interval $(r_2,\infty)$, the Gibbs free energy is negative and decreasing with respect to $r_h$. Hence, the black hole with $r_h>r_2$ is globally stable. Note that, the Gibbs free energy vanishes at the points $r_1$ and $r_2$. These points refer to the horizons of those black holes which can experience the Hawking-Page phase transition. Thermodynamically stable black holes with horizon radii greater than both $r_1$ and $r_2$ are in the radiation phase. The plot of Gibbs energy in Fig. \ref{skr5B} shows that the black holes of massive gravity's rainbow with greater values of $m_g(\varepsilon)$ are more globally stable than the objects associated with smaller values of this parameter. Similarly, Fig. \ref{skr5C} shows that the parameter $\beta(\varepsilon)$ does not directly produce any significant effect on global stability. It should be noted that the nonlinearity parameter has an indirect effect on the Gibbs energy that comes from the horizon radius through Eq. (\ref{24}). Similarly, Fig. \ref{skr5D} demonstrates the behaviour of Gibbs energy for different values of charge $Q$. It can be seen that the smaller black holes associated with the lesser magnitude of charge $Q$ are more globally stable than the objects of the same size and with a greater magnitude of charge. However, when the size of the black hole increases, all the magnetized black holes described by Eq. (\ref{17}) are enjoying global stability.   
\section{Summary and conclusion} 

In this work, I assumed gravity's rainbow for the inclusion of UV limit in Einstein-massive gravity. It is figured out to find the energy-dependent anti-de Sitter black hole solution in massive gravity coupled to NLED. In general $d$-dimensional spacetime, the model of double-logarithmic electrodynamics has been used as the matter source for gravity. It is demonstrated that the proposed solution (\ref{17}) can be understood as black holes with one or two horizons, or even naked singularities. It is important to note that it also describes the magnetized black holes of gravity's rainbow when $m_g(\varepsilon)$ is equal to zero. Furthermore, for $\beta(\varepsilon)$ approaches to zero, one can get the black hole solution with a standard Maxwell source. However, by using $m_g(\varepsilon)=\Lambda(\varepsilon)=0$, $g(\varepsilon)=h(\varepsilon)=1$ and $\beta(\varepsilon)\rightarrow 0$ in Eq. (\ref{17}), one can find the analogous magnetic version of the RN black hole solution. 

The mathematical expressions for different thermodynamic quantities associated with (\ref{17}) are also derived. It is concluded that the graviton's mass, the rainbow functions, and the nonlinearity parameter $\beta(\varepsilon)$ all affect these quantities. Additionally, the generalized first law and Smarr's formula corresponding to the metric function (\ref{17}) are also constructed. Plots of the temperature and heat capacity are made and the local thermodynamic stability is investigated. The black holes with such horizon radii $r_h$ are locally stable for which the temperature and heat capacity are nonnegative and otherwise unstable. It is demonstrated that in the higher spacetime dimensions, the stable black holes' critical horizon radii are larger. It is to be noted that the black holes whose horizon radii are less than their associated critical value are unstable. Thus, these critical horizon radii coincide with the radii of extreme black holes. Additionally, it is concluded that the greater region of local stability is provided by the smaller values of the rainbow functions. On the other side, when the rainbow functions grow, the region corresponding to the local thermodynamic stability shrinks. Furthermore, as the values of $m_g(\varepsilon)$ grow, the domain of $r_h$ linked with the positivity of heat capacity expands. The regions of the local thermodynamic stability are also highlighted for various values of the nonlinearity parameter and magnetic charge. It is explored that the local stability of black holes is enhanced by a higher magnitude of the nonlinearity parameter. Moreover, the black holes are becoming more locally stable when the magnetic charge decreases. The intervals of $r_h$ related to global stability are discovered, and the precise formula for the Gibbs free energy is also determined. It is demonstrated that the global thermodynamic stability can be significantly influenced by the massive graviton, dimensionality parameter, magnetic charge, and rainbow functions.

The investigation related to critical behaviour, quasi-normal modes \cite{gf}, Hawking radiation \cite{new, new1} and thermal fluctuations of the black holes obtained in this paper could be significant. Likewise, one can also investigate the effects of double-logarithmic electrodynamics on the axially symmetric and cylindrical black holes of massive gravity's rainbow. I have left these ideas for my future investigations.
\newline \newline
\textbf{Data availability statement}
This manuscript has no associated data or the data will not be deposited. \newline
\newline
\textbf{Conflict of Interest}
The author declares no competing interests.


\begin{thebibliography}{150}
	
\bibitem{1} D. Lovelock, J. Math. Phys. \textbf{12} (1971) 498.

\bibitem{2} D. Lovelock, J. Math. Phys. \textbf{13} (1972) 874.

\bibitem{3} C. Brans and R. H. Dicke, Phys, Rev. \textbf{124} (1961) 925.
 
\bibitem{4} R. G. Cai and Y. S. Myung, Phys. Rev. D \textbf{56} (1997) 3466.

\bibitem{5} P. Brax and C. Van de Bruck, Class. Quantum Grav. \textbf{20} (2003) R201.

\bibitem{6} L. A. Gargely, Phys. Rev. D \textbf{74} (2006) 024002.

\bibitem{7} J. C. de Souza and V. Faroni, Class. Quantum Grav. \textbf{24} (2007) 3637.

\bibitem{8} G. Cognola, E. Elizalde, S. Nojiri, S. Odintsov, L. Sebastiani and S. Zerbini, Phys. Rev. D \textbf{77} (2008) 046009.

\bibitem{9} T. P. Sotiriou and V. Faroni, Rev. Mod. Phys. \textbf{82} (2010) 451.

\bibitem{10} S. Nojiri, S. D. Odintsov and V. K. Oikonomou, Phys. Rept. \textbf{692} (2017) 1.

\bibitem{11} J. Magueijo and L. Smolin, Phys. Rev. Lett. \textbf{88} (2002) 190403.

\bibitem{12} J. Magueijo and L. Smolin, Class. Quantum Grav. \textbf{21} (2004) 1725.

\bibitem{13} A. Einstein, Ann. Phys. (Berlin) \textbf{17} (1905) 891.

\bibitem{14} G. Amelino-Camelia, Phys. Lett. B \textbf{510} (2001) 255.

\bibitem{15} G. Amelino-Camelia and J. Kowalski-Glikmann, Phys. Lett. B \textbf{522} (2001) 133.

\bibitem{16} J. Kowalski-Glikmann, Phys. Lett. A \textbf{286} (2001) 391.

\bibitem{17} J. J. Peng and S. Q. Wu, Gen. Relativ. Gravit. \textbf{40} (2008) 2619.

\bibitem{18} A. F. Ali and M. M. Khalil, Eur. Phys. Lett. \textbf{110} (2015) 20009.

\bibitem{19} A. F. Ali, M. Faizal and B. Majumder, Eur. Phys. Lett. \textbf{109} (2015) 20001.

\bibitem{20} Y. Gim and W. Kim, J. Cosmol. Astropart. Phys. \textbf{05} (2015) 002.

\bibitem{21} Y. Ling, X. Li and H. B. Zhang, Mod. Phys. Lett. A \textbf{22} (2007) 2749.

\bibitem{22} H. Li, Y. Ling and X. Han, Class. Quantum Grav. \textbf{26} (2009) 065004.

\bibitem{23} A. F. Ali, Phys. Rev. D \textbf{89} (2014) 104040.

\bibitem{24} A. Awad, A. F. Ali and B. Majumder, J. Cosmol. Astropart. Phys. \textbf{10} (2013) 052.

\bibitem{25} S. H. Hendi, M. Momennia, B. Eslam Panah, M. Faizal, Astrophys. J. \textbf{827} (2016) 153.

\bibitem{26} M. Khodadi, K. Nozari, H. R. Sepangi, Gen. Relativ. Gravit. \textbf{48} (2016) 166.

\bibitem{27} M. Khodadi, Y. Heydarzade, K. Nozari, F. Darabi, Eur. Phys. J. C \textbf{75} (2015) 590.

\bibitem{28} S. H. Hendi, G. H. Bordbar, B. Eslam Panah, S. Panahiyan, J. Cosmol. Astropart. Phys. \textbf{09} (2016) 013. 

\bibitem{29} B. Eslam Panah, G. H. Bordbar, S. H. Hendi, R. Ruffini, Z. Rezaei, R. Moradi, Astrophys. J. \textbf{848} (2017) 24.

\bibitem{30} R. Garattini, G. Mandanici, Eur. Phys. J. C \textbf{77} (2017) 57.
 
\bibitem{31} R. Garattini, E. N. Saridakis, Eur. Phys. J. C \textbf{75} (2015) 343.

\bibitem{32} S. Alsaleh, Eur. Phys. J. Plus \textbf{132} (2017) 181.

\bibitem{33} C. Leiva, J. Saavedra, J. Villanueva, Mod. Phys. Lett. A \textbf{24} (2009) 1443.

\bibitem{34} S. H. Hendi, S. Panahiyan, B. Eslam Panah and M. Momennia, Eur. Phys. J. C \textbf{76} (2016) 150.

\bibitem{36} S. H. Hendi, S. Panahiyan, B. Eslam Panah, M. Faizal and M. Momennia, Phys. Rev. D \textbf{94} (2016) 024028.

\bibitem{37} S. H. Hendi, B. Eslam Panah and S. Panahiyan, Phys. Lett. B \textbf{769} (2017) 191.

\bibitem{38} R. Garattini, J. Cosmol. Astropart. Phys. \textbf{06} (2013) 017.

\bibitem{39} S. H. Hendi, S. Panahiyan, B. Eslam Panah and M. Momennia, Adv. High energy Phys. \textbf{2016} (2016) 9813582.	

\bibitem{1a} S. Deser, R. Jackiw and G. Hooft, Ann. Phys. \textbf{152} (1984) 220.

\bibitem{2a} M. Fierz, Helv. Phys. Acta \textbf{12} (1939) 3; M. Fierz and W. Pauli, Proc. Roy. Soc. Lond. A \textbf{173} (1939) 211.

\bibitem{3a} D. G. Boulware and S. Deser, Phys. Rev. D \textbf{6} (1972) 3368.

\bibitem{4a} S. F. Hassan and R. A. Rosen, Phys. Rev. Lett. \textbf{108} (2012) 041101; S. F. Hassan, R. A. Rosen and A. Schmidt-May, J. High Energy Phys. \textbf{02} (2012) 026.

\bibitem{5a} P. Minjoon, Class. Quantum Grav. \textbf{28} (2011) 105012.

\bibitem{6a} C. de Rham and G. Gabadadze, Phys. Rev. D \textbf{82} (2010) 044020; C. de Rham, G. Gabadadze and A. J. Tolley, Phys. Rev. Lett. \textbf{106} (2011) 231101.

\bibitem{7a} K. Hinterbichler, Rev. Mod. Phys. \textbf{84} (2012) 671.

\bibitem{8a} C. de Rham, Massive gravity, Living Rev. Rel. \textbf{17} (2014) 7.

\bibitem{9a} E. Babichev, L. Marzola, M. Raidal, A. Schmidt-May, F. Urban, H. Veerme, and M. Von Strauss, Phys. Rev. D \textbf{94} (2016) 084055.

\bibitem{10a} E. Babichev, L. Marzola, M. Raidal, A. Schmidt-May, F. Urban, H. Veerme, and M. Von Strauss, J. Cosmol. Astropart. Phys. \textbf{09} (2016) 016.

\bibitem{11a} Y. Akrami, T. S. Koivisto and M. Sandstad, J. High Energy Phys. \textbf{03} (2013) 99.

\bibitem{12a} Y. Akrami, S. F. Hassan, F. Knnig, A. Schmidt-May, and A. R. Solomon, Phys. Lett. B \textbf{748} (2015) 37.

\bibitem{13a} D. Vegh,  Holography without translational symmetry, arXiv:1301.0537.

\bibitem{14a} S. F. Hassan and R. A. Rosen, J. High Energy Phys. \textbf{07} (2011) 009.

\bibitem{15a} R. A. Davison, Phys. Rev. D \textbf{88} (2013) 086003; M. Blake and D. Tong, Phys. Rev. D \textbf{88} (2013) 106004; R. A. Davison, K. Schalm and J. Zaanen, Phys. Rev. B \textbf{89}, (2014) 245116.

\bibitem{16a} M. Baggioli and O. Pujolas, Phys. Rev. Lett. \textbf{114}, (2015) 251602.

\bibitem{17a} S. F. Hassan, R. A. Rosen and A. Schmidt-May, J. High Energy Phys. \textbf{02} (2012) 026.

\bibitem{18a} L. Alberte and A. Khmelnitsky, Phys. Rev. D \textbf{88} (2013) 064053.

\bibitem{19a} L. Alberte and A. Khmelnitsky, Phys. Rev. D \textbf{91} (2015) 046006.

\bibitem{20a} H. Zhang and X-Z. Li, Phys. Rev. D \textbf{93} (2016) 124039.

\bibitem{21a} T. Q. Do, Phys. Rev. D \textbf{93} (2016) 104003.

\bibitem{22a} T. Q. Do, Phys. Rev. D \textbf{94} (2016) 044022.

\bibitem{23a} G. D'Amico, C. de Rham, S. Dubovsky, G. Gabadadze, D. Pirtskhalava and A. J. Tolley, Phys. Rev. D \textbf{84} (2011) 124046.

\bibitem{24a} A. H. Chamseddine and M. S. Volkov, Phys. Lett. B \textbf{704} (2011) 652.

\bibitem{25a} C. de Rham, J. T. Deskins, A. J. Tolley and S. Y. Zhou, Rev. Mod. Phys. \textbf{89} (2017) 025004.

\bibitem{26a} K. Koyama, G. Niz and G. Tasinato, Phys. Rev. Lett. \textbf{107} (2011) 131101.

\bibitem{27a} T. M. Nieuwenhuizen, Phys. Rev. D \textbf{84} (2011) 024038.

\bibitem{28a} A. Gruzinov and M. Mirbabayi, Phys. Rev. D \textbf{84} (2011) 124019.

\bibitem{29a} L. Berezhiani, G. Chkareuli, C. de Rham, G. Gabadadze and A. J. Tolley, Phys. Rev. D \textbf{85} (2012) 044024.

\bibitem{30a} R. A. Rosen, J. High Energy Phys. \textbf{10} (2017) 206.

\bibitem{31a} A. Dehghani, S. H. Hendi and R. B. Mann, Phys. Rev. D \textbf{101} (2020) 084026.

\bibitem{81} M. Born and L. Infeld, Proc. R. Soc. Lond. A \textbf{144} (1934) 425.

\bibitem{82} E. S. Fradkin and A. Tseytlin, Phys. Lett. B \textbf{163} (1985) 123.

\bibitem{83} A. Tseytlin, Nucl. Phys. B \textbf{276} (1986) 391.

\bibitem{84} D. M. Gitman and A. E. Shabad, Eur. Phys. J. C \textbf{74} (2014) 3186.

\bibitem{85} S. I. Kruglov, Eur. Phys. J. C \textbf{75} (2015) 88.

\bibitem{86} S. I. Kruglov, Ann. Phys. \textbf{353} (2015) 299.

\bibitem{87} S. I. Kruglov, Ann. Phys. \textbf{527} (2015) 397.

\bibitem{88} I. Gullu and S. H. Mazharimousavi, Phys. Scr. \textbf{96} (2021) 045217.

\bibitem{89} H. P. de Oliveira, Class Quantum Grav. \textbf{11} (1994) 1469.

\bibitem{90} W. Heisenberg and H. Euler, Z. Phys. \textbf{98} (1936) 714.

\bibitem{91} E. Ayon-Beato and A. Garcia, Phys. Rev. Lett. \textbf{80} (1998) 5056.

\bibitem{92} L. Balart and E. C. Vagenas, Phys. Rev. D \textbf{90} (2014) 124045.

\bibitem{93} N. Breton, Phys. Rev. D \textbf{67} (2003) 124004.

\bibitem{94} I. Dymnikova, Class. Quantum Grav. \textbf{21} (2004) 4417.

\bibitem{95} S. I. Kruglov, Int. J. Geom. Meth. Mod. Phys. \textbf{12} (2015) 1550073.

\bibitem{98} S. Fernando and D. Krug, Gen. Relativ. Gravit. \textbf{35}, 129 (2003).

\bibitem{99} T. K. Dey, Phys. Lett. B \textbf{595}, 484 (2004).

\bibitem{100} R. G. Cai, D. W. Pang and A. Wang, Phys. Rev. D \textbf{70} (2004) 124034.

\bibitem{102} M. Aiello, R. Ferraro and G. Giribet, Phys. Rev. D {70} (2004) 104014.

\bibitem{103} J. Diaz-Alonso and D. Rubiera-Garcia, Phys. Rev. D \textbf{81} (2010) 064021.

\bibitem{107} H. H. Soleng, Phys. Rev. D \textbf{52} (1995) 6178.

\bibitem{108} H. Yajima and T. Tamaki, Phys. Rev. D \textbf{63} (2001) 064007.

\bibitem{109} E. Ayon-Beato and A. Garcia, Phys. Lett. B \textbf{464} (1999) 25.

\bibitem{113} K. A. Bronnikov, Phys. Rev. D \textbf{63} (2001) 044005.

\bibitem{114} M. Hassaine and C. Martinez, Phys. Rev. D \textbf{75} (2007) 027502.

\bibitem{116} H. A. Gonzalez, M. Hassaine and C. Martinez, Phys. Rev. D \textbf{80} (2009) 104008.

\bibitem{117} S. H. Mazharimousavi, M. Halilsoy and O. Gurtug, Class. Quantum Grav. \textbf{27} (2010) 205022.

\bibitem{118} M. H. Dehghani and H. R. R. Sedehi, Phys. Rev. D \textbf{74} (2006) 124018.

\bibitem{119} M. H. Dehghani, S. H. Hendi, A. Sheykhi and H. R. Sedehi, J. Cosmol. Astropart. Phys. \textbf{02} (2007) 020.

\bibitem{120} S. H. Hendi, Phys. Rev. D \textbf{82} (2010) 064040.

\bibitem{121} S. I. Kruglov, Ann. Phys. \textbf{383} (2017) 550.

\bibitem{122} S. I. Kruglov, Ann. Phys. \textbf{378} (2017) 59.

\bibitem{123} S. I. Kruglov, Int. J. Mod. Phys. A \textbf{33} (2018) 1850023.

\bibitem{124} S. I. Kruglov, Int. J. Mod. Phys. A \textbf{32} (2017) 1750147.

\bibitem{125} S. I. Kruglov, Ann. Phys. (Berlin) \textbf{529} (2017) 1700073.

\bibitem{126} A. Ali and K. Saifullah, Ann. Phys. \textbf{437} (2022) 168726.

\bibitem{126b} A. Ali, Int. J. Geom. Meth. Mod. Phys. \textbf{18} (2021) 2150184.

\bibitem{127} S. I. Kruglov, Phys. Lett. A \textbf{379} (2015) 623.

\bibitem{127a} Y. F. Cai, D. A. Easson, C. Gao and E. N. Saridakis, Phys. Rev. D \textbf{87} (2013) 064001.

\bibitem{127b} E. Babichev and A. Fabbri, J. High Energy Phys. \textbf{07} (2014) 016.

\bibitem{127c} R. G. Cai, Y. P. Hu, Q. Y. Pan and Y. L. Zhang, Phys. Rev. D \textbf{91} (2015) 024032.

\bibitem{127d} J. Xu, L. M. Cao and Y. P. Hu, Phys. Rev. D \textbf{91} (2015) 124033.

\bibitem{127f} A. Ali and K. Saifullah, Eur. Phys. J. C \textbf{82} (2022) 131.

\bibitem{127f1} S. H. Hendi, B. Eslam Panah and S. Panahiyan, Fortschr. Phys. \textbf{66} (2018) 1800005.

\bibitem{127f2} S. H. Hendi, B. Eslam Panah and S. Panahiyan, J. High Energy Phys. \textbf{11} (2015) 157.

\bibitem{127g} A. Ali and K. Saifullah, Phys. Lett. B \textbf{792} (2019) 276. 

\bibitem{127h} K. Meng and D. B. Yang, Phys. Lett. B \textbf{780} (2018) 363.

\bibitem{127i} A. Ali and K. Saifullah, Phys. Rev. D \textbf{99} (2019) 124052.

\bibitem{127j} I. Gullu and S. H. Mazharimousavi, Phys. Scr. \textbf{96} (2021) 095213.

\bibitem{127k} A. Ali, Eur. Phys. J. Plus \textbf{137} (2022) 108.

\bibitem{127r1} S. I. Kruglov, Grav. Cosmol. \textbf{25} (2019) 190-195.

\bibitem{127r2} S. Kanzi, S. H. Mazharimousavi and I. Sakalli, Ann. Phys. \textbf{422} (2020) 168301.

\bibitem{128a} B. Eslam Panah and S. H. Hendi, Phys. Lett. B \textbf{684} (2010) 77.

\bibitem{my1} A. Ali and K. Saifullah, Rotating black branes in Lovelock gravity with double-logarithmic electrodynamics, Annals of Physics, 2022 (in press).

\bibitem{128b} S. H. Hendi, B. Eslam Panah, S. Panahiyan and M. Hassaine, Phys. Rev. D \textbf{98} (2018) 084006.

\bibitem{128c} S. H. Hendi, B. Eslam Panah, S. Panahiyan and M. Momennia, Eur. Phys. J. C \textbf{77} (2017) 647.

\bibitem{128d} A. Sheykhi, M. H. Dehghani, Adv. High Energy. Phys. \textbf{2016} (2016) 3265968.

\bibitem{128E} I. Sakalli, M. Halilsoy and H. Pasaoglu, Int. J. Theor. Phys. \textbf{50} (2011) 3212-3224.

\bibitem{128} R. Garcia-Salcedo and N. Breton, Int. J. Mod. Phys. A \textbf{15} (2000) 4341.

\bibitem{129} C. S. Camara, M. R. de Garcia Maia, J. C. Carvalho and J. A. S. Lima, Phys. Rev. D \textbf{69} (2004) 123504.

\bibitem{130} E. Elizalde, J. E. Lidsey, S. Nojiri and S. D. Odintsov, Phys. Lett. B \textbf{574} (2003) 1. 

\bibitem{131} M. Novello, S. E. Perez Berglia and J. M. Salim, Phys. Rev. D \textbf{69} (2004) 127301.

\bibitem{132} M. Novello, E. Goulart, J. M. Salim and S. E. Perez Berglia, Class. Quantum Grav. \textbf{24} (2007) 3021.

\bibitem{133} D. N. Vollick, Phys. Rev. D \textbf{78} (2008) 063524.

\bibitem{134} S. I. Kruglov, Phys. Rev. D \textbf{92} (2015) 123523.

\bibitem{135} S. I. Kruglov, Int. J. Mod. Phys. D \textbf{25} (2016) 1640002.

\bibitem{138} L. Smarr, Phys. Rev. Lett. \textbf{30} (1973) 71.

\bibitem{139} N. Breton, Gen. Relativ. Gravit. \textbf{37} (2005) 643.

\bibitem{141} Y. Zhang and S. Gao, Class. Quantum Grav. \textbf{35} (2018) 145007.

\bibitem{143} P. Wang, H. Wu and H. Yang, Eur. Phys. J. C \textbf{79} (2019) 572.

\bibitem{144} S. Hendi, R. B. Mann, S. Panahiyan and B. Eslam Panah, Phys. Rev. D \textbf{95} (2017) 021501.

\bibitem{144a} S. Fernando, Int. J. Mod. Phys. D \textbf{26} (2017) 1750071.

\bibitem{144b} S. G. Ghosh, Int. J. Mod. Phys. D \textbf{21} (2012) 1250022.

\bibitem{144c} Md Sabir Ali and S. G. Ghosh, Phys. Rev. D \textbf{98} (2018) 084025.

\bibitem{144d} A. Kumar, D. V. Singh and S. G. Ghosh, Eur. Phys. J. C \textbf{79} (2019) 275.

\bibitem{145} L. Brewin, Gen. Relativ. Gravit. \textbf{39} (2007) 521.  

\bibitem{146} S. W. Hawking, Commun. Math. Phys. \textbf{43} (1975) 199.

\bibitem{147} J. D. Bekenstein, Phys. Rev. D \textbf{7} (1973) 2333.

\bibitem{148} C. J. Hunter, Phys. Rev. D \textbf{59} (1999) 024009.

\bibitem{149} M. Dehghani and B. Badpa, Prog. Theor. Phys. \textbf{2020} (2020) 033E03.

\bibitem{gf} I. Sakalli and S. Kanzi, Turk. J. Phys. \textbf{46} (2022) 51-103.

\bibitem{new} I. Sakalli and A. Ovgun, Eur. Phys. Lett. \textbf{110} (2015) 10008.

\bibitem{new1} I. Sakalli, M. Halilsoy and H. Pasaoglu, Astrophys. Space Sci. \textbf{340} (2012) 155-160.



\end{thebibliography}
\end{document}